# Enhancements to A Lightweight RFID Authentication Protocol

Xiaowen Zhang[1], Zhanyang Zhang[1], Xinzhou Wei[2]

[1]Dept. of Computer Science, College of Staten Island / CUNY, Staten Island, NY 10314
[2]Dept. of ETET, New York City College of Technology / CUNY, Brooklyn, NY 11201

**Abstract**

Vajda and Buttyan (VB) proposed a set of five lightweight RFID authentication protocols. Defend, Fu, and Juels (DFJ) did cryptanalysis on two of them – XOR and SUBSET. To the XOR protocol, DFJ proposed repeated keys attack and nibble attack. In this paper, we identify the vulnerability existed in the original VB's successive session key permutation algorithm. We propose three enhancements to prevent DFJ's attacks and make XOR protocol stronger without introducing extra resource cost.

## 1 INTRODUCTION

Along with the massive deployment of Radio Frequency Identification (RFID) systems in variety of applications, many security issues and privacy concerns have been brought up. Some consumer right protection organizations, like CASPIAN (Consumers Against Supermarket Privacy Invasion and Numbering), are against the use of RFID [2].

In general an RFID system consists of three kinds of components: many (thousands to millions) RFID tags (or transponders), several RFID readers (or interrogators), and a few backend computer servers. A RFID tag is a tiny microchip equipped with radio frequency antenna. it is capable of emitting the identification and other related data for the tagged item. A reader is another electronic device located between tags and backend server. A reader gets information from or sends information to the tag. It communicates with (updates) the backend server. A backend server runs applications software, hosts databases, processes tag information received from a reader. A server acts as a gateway. It communicates (through wireless or wire) with readers on one end and with the enterprise network (the Internet) infrastructure on the other end. The wireless communication links between tags and readers are considered the most vulnerable to security and privacy threats. As documented in many literature [1, 7, 13], RFID and security experts have devoted a lot of efforts to address these threats. Among those efforts, new RFID authentication protocols and analysis are active areas of research [3, 5, 6, 8, 9, 10, 12].

Adding security features to low-cost RFID tags is a daunting and challenging task because these tags are extremely resource limited and cannot afford for strong cryptographic algorithms. Practical RFID authentication protocols should have the following characteristics: lightweight, anonymity (un-traceability), mutual authentication.

Vajda and Buttyan (VB) [14] proposed a set of five lightweight RFID authentication protocols and also gave a brief analysis. Each one of the protocols is extremely lightweight in terms of resources required, and is considered suitable for resource limited devices, like RFID tags.

Defend, Fu, and Juels (DFJ) [4] did cryptanalysis to two of them – XOR and SUBSET. DFJ proposed repeated keys attack and nibble attack to compromise the XOR protocol. In this paper, we identify the vulnerability existed in the original VB's successive session key permutation algorithm. We propose three enhancements, removing bad shuffles, hopping the runs, and authenticating mutually, to prevent DFJ's attacks and make XOR protocol stronger without introducing extra resource cost.

## 2 ORIGINAL XOR PROTOCOL AND REPEATED KEYS ATTACK

The original XOR protocol by VB [14] is a challenge-response protocol. (see Figure 1). Providing the following assumptions, (1) the readers and tags share a piece of secret key $k^{(0)}$ initially, (2) both reader and tag are capable of calculating a permutation $\prod$ (given soon), (3) reader and tag maintain a synchronized counter $i$ to indicate the current run of authentication, the challenge-response process at the ith run can be described as:

Reader $\rightarrow$ Tag: $a^{(i)} = x^{(i)} \oplus k^{(i)}$

// Reader picks a random number x(i), calculates $k^{(i)}$, then sends a challenge $a^{(i)} = x^{(i)} \oplus k^{(i)}$ to Tag.

Tag $\rightarrow$ Reader: $b^{(i)} = x^{(i)} \oplus k^{(0)}$

// Tag calculates $k^{(i)}$, extracts the challenge $x^{(i)}$ by $k^{(i)} \oplus a^{(i)}$, then send a response $b^{(i)} = x^{(i)} \oplus k^{(0)}$ to Reader. Then the Reader verifies the Tag, because only the Tag knows $k^{(0)}$.

Here $k^{(i)} = \prod(k^{(i-1)})$, and $\prod: \{0, 1\}^n \rightarrow \{0, 1\}^n$ is a permutation starting from the initial secret key $k^{(0)}$. That is $k^{(1)} = \prod(k^{(0)})$, $k^{(2)} = \prod(k^{(1)})$, …, $k^{(i-1)} = \prod(k^{(i-2)})$, $k^{(i)} = \prod(k^{(i-1)})$, …. Because $x^{(i)}$ is random, so are $a^{(i)} = x^{(i)} \oplus k^{(i)}$ and $b^{(i)} = x^{(i)} \oplus k^{(0)}$. If the $x^{(i)}$ is truly random, no information about the secret $k^{(0)}$ are revealed from the communication.

Suppose n = 128 bit as key length, the steps of the permutation $\prod$ is given as follows:

- Step-1: In run (i-1), the session key $k^{(i-1)}$ is split into 16 bytes, then cut each byte into two nibbles of 4-bit each. Then concatenate all left nibbles $k_{0,L}^{(i-1)}$, $k_{1,L}^{(i-1)}$, …, $k_{15,L}^{(i-1)}$ to form $k_L^{(i-1)}$, concatenate all right nibbles $k_{0,R}^{(i-1)}$, $k_{1,R}^{(i-1)}$, …, $k_{15,R}^{(i-1)}$ to form $k_R^{(i-1)}$.
- Step-2: For the run (i) the right half key $k_R^{(i)}$ is a permutation of $k_R^{(i-1)}$ controlled by $k_L^{(i-1)}$: i.e., swapping the 0-th and the $k_{0,L}^{(i-1)}$-th, the 1-st and the $k_{1,L}^{(i-1)}$-th, …, the 15-th and the $k_{15,L}^{(i-1)}$-th nibbles of $k_R^{(i-1)}$.
- Step-3: The left half key $k_L^{(i)}$ for round (i) is a permutation of $k_L^{(i-1)}$ controlled by $k_R^{(i-1)}$ in the similar nibble swaps.
- Step-4: Finally the next run session key $k^{(i)}$ is the obtained from rearranging (interleaving) the half bytes of $k_L^{(i)}$ and $k_L^{(i)}$, i.e., $k^{(i)} = k_{0,L}^{(i)} | k_{0,R}^{(i)} | k_{1,L}^{(i)} | k_{1,R}^{(i)} | \cdots | k_{15,L}^{(i)} | k_{15,R}^{(i)}$, here "|" represents concatenation.

**Observation:** The problem we see here is the step-4 of the permutation. This step adds a perfect shuffle to $\prod$. No matter what we do out-shuffle $k^{(i)} = k_{0,L}^{(i)} | k_{0,R}^{(i)} | k_{1,L}^{(i)} | k_{1,R}^{(i)} | \cdots | k_{15,L}^{(i)} | k_{15,R}^{(i)}$ or in-shuffle $k^{(i)} = k_{0,R}^{(i)} | k_{0,L}^{(i)} | k_{1,R}^{(i)} | k_{1,L}^{(i)} | \cdots | k_{15,R}^{(i)} | k_{15,L}^{(i)}$, after some number of shuffles the sequence will return to the original order [11]. That is the culprit why this permutation suffers from short cycles, as DFJ [4] identified, and makes the protocol vulnerable to their repeated keys attack.

Based on experiments by DFJ [4] given a initial key $k^{(0)}$ the successive session keys $k^{(1)}$, $k^{(2)}$, …, $k^{(i)}$, …, $k^{(10000)}$, which are generated by the permutation $\prod$, cycle after an average of 68 sessions. They also found that about 32% of session keys have cycle 1, and all of tested session keys eventually repeat and only one thousandth of keys have the maximum cycle of 36. Let c represents a cycle, then $k^{(i)} = k^{(i+c)}$.

Suppose $k^{(i)} = k^{(i-2)}$, the adversary model for the repeated keys attack [4] is shown in Figure 1. The eavesdropper Eve is able to form a valid response without knowing $k^{(0)}$ or $x^{(i)}$, therefore she can impersonate a valid tag.

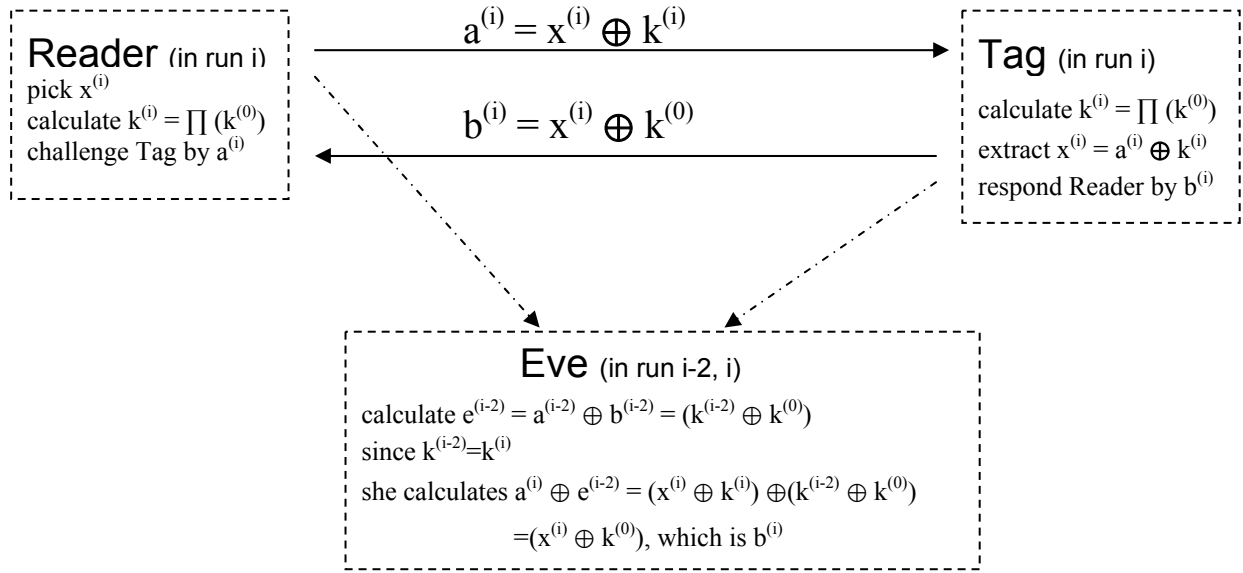

Figure 1. Adversary model for repeated keys attack.

## 3  ENHANCEMENTS TO XOR PROTOCOL

In this section we propose three enhancements to the original XOR protocol, namely Removing bad shuffles, Hopping the runs, and Authenticating mutually.

## 3.1 Enhancement I – Removing Bad Shuffles

In the step-4 of the original VB's permutation algorithm, we don't interleaving nibbles of $k_L^{(i)}$ and $k_R^{(i)}$ to create $k^{(i)}$. Instead, we just simply concatenate $k_L^{(i)}$ and $k_R^{(i)}$ to form $k^{(i)}$, i.e., $k^{(i)} = k_L^{(i)} | k_R^{(i)}$. In this way we remove that out-shuffles from the permutation, the short cycles disappear, and the DFJ's repeated key attack is prevented. And we argue that without the out-shuffle step, the permutation $\prod$ is a Knuth shuffle [15][1], i.e., an algorithm for generating a random permutation of a finite set. In the case of 128-bit key length, there are two finite sets with 16 nibbles (0 ~ F in hexadecimal) each. The algorithm $\prod$ to the left and right nibble sets ($k_L^{(i)}$ and $k_R^{(i)}$) will give 16! permutations each, in theory there are 16!*16! (about $2^{89}$) in total.

Here is an example to illustrate how the permutation $\prod$ without out-shuffles works. We use a pseudo-random key generation program to create a 128-bit $k^{(0)}$ in hexadecimal as: C3 47 3F BB 8D B4 C1 E0 5F 4C 2D 8B 2B A6 BD 98, then split it into left and right nibble sets as
$k_L^{(0)}$: C 4 3 B 8 B C E 5 4 2 8 2 A B 9,
$k_R^{(0)}$: 3 7 F B D 4 1 0 F C D B B 6 D 8.
Then under the control of $k_R^{(0)} / k_L^{(0)}$ we permute $k_L^{(0)} / k_R^{(0)}$ to obtain $k_L^{(1)} / k_R^{(1)}$ as follows:
$k_L^{(1)}$: 4C 98 BA 2B 52 84 CB E3 and
$k_R^{(1)}$: BD 1B C7 3D 48 60 DB FF.
$k^{(1)}$ is the concatenation of $k_L^{(1)}$ and $k_R^{(1)}$ as:
$k^{(1)}$: 4C 98 BA 2B 52 84 CB E3 BD 1B C7 3D 48 60 DB FF.

The second run starts from $k^{(1)}$: 4C 98 BA 2B 52 84 CB E3 BD 1B C7 3D 48 60 DB FF, again we split it into left and right nibble sets as
$k_L^{(1)}$: 4 9 B 2 5 8 C E B 1 C 3 4 6 D F,
$k_R^{(1)}$: C 8 A B 2 4 B 3 D B 7 D 8 0 B F.
Then controlled by $k_R^{(1)} / k_L^{(1)}$ permute $k_L^{(1)} / k_R^{(1)}$ to obtain $k_L^{(2)} / k_R^{(2)}$ as follows:
$k_L^{(2)}$: 1B 5E 8C 2B 4C 3D 64 9F and
$k_R^{(2)}$: 28 BC 7D 0B AB BD 43 8F.
$k^{(1)}$ is the concatenation of $k_L^{(1)}$ and $k_R^{(1)}$ as:
$k^{(2)}$: 1B 5E 8C 2B 4C 3D 64 9F 28 BC 7D 0B AB BD 43 8F.

And so on so forth, we can get $k^{(3)}$, $k^{(4)}$, $k^{(5)}$, …
$k^{(3)}$: 04 A5 6B 88 24 3B 27 19 FD CB 38 4D FB BD CE CB,
$k^{(4)}$: 23 4F 06 1C 3A BC 82 CF 49 B5 DD 48 BB BD B7 8E,
$k^{(5)}$: 18 0B D4 8B 42 BB B3 C4 3C 6F 58 9B EC D7 2F DA.

---

[1] "Knuth shuffle is to the identity permutation or any other any permutation, then go through the positions 1 through n−1, and for each position *i* swap the element currently there with an arbitrarily chosen element from positions *i* through n, inclusive. It's easy to verify that any permutation of n elements will be produced by this algorithm with probability exactly 1/*n*!, thus yielding a uniform distribution over all such permutations."

**Observation:** the permutation ∏ does not create the new nibbles (hexadecimal symbols), instead it just move all existing nibbles around by each run. If an ideal pseudo-random key generator is used during the initial key generation, within $k^{(0)}$ nibbles (0 ~ F) should uniformly distributed and unbiased. Theoretically, the random permutation ∏ can guarantee 16!*16! permutations for a 128-bit sequence. However, if the distribution of 16 nibbles is not uniform (as the above example shows the frequencies of the 16 hex symbols are: 0(1), 1(1), 2(2), 3(2), 4(3), 5(1), 6(1), 7(1), 8(3), 9(1), A(1), B(6), C(3), D(3), E(1), F(2) ), the total number of permutations is less than 16!*16!, but still a huge number.

In practice, before installation of a $k^{(0)}$'s to a tag, this $k^{(0)}$ should be tested to make sure this $k^{(0)}$ can be used to generate enough number (5*16! is already big enough) of session keys without repetition. This procedure is used to eliminate weak keys. If key length is 128-bit, there are in total $2^{128}$ $k^{(0)}$'s. Even only one good strong key among every one hundred keys, still the number of strong keys is about $2^{121}$, it's huge.

For curiosity we carried out an experiment, similarly as in [4], with a 128-bit key length. We generated 1000 different $k^{(0)}$'s, and from each $k^{(0)}$ we permutated 10000 times to get session keys $k^{(1)}, k^{(2)}, ..., k^{(i-1)}, k^{(i)}, k^{(i+1)}, ..., k^{(10000)}$ to put them into a file. We obtained 1000 such files with each containing 10000 session keys. As far as our experiment concerns, we did not find a repeat session key within these 1000 files. So the repeated keys attack is prevented.

### 3.2 Enhancement II – Hopping the Runs

The reason of making the session keys hop is that the next session key does not have to be the one immediately successor of the current session key. This makes the nibble attack [4] much harder, if it's not impossible.

Here is how the nibble attack [4] works. Starts from run *i*, the attacker Eve builds a table over the following number of runs. Two columns are the challenges and responses between Reader and Tag, i.e., a's and b's. The next column is the xor'ed result the previous two columns, i.e., $a^{(i)} \oplus b^{(i)} = k^{(i)} \oplus k^{(0)}$. The last column is the xor'ed result of two consecutive rows from the fourth column, which will give us the xor'ed result of two consecutive session keys. As observed when a nibble of this last column becomes 0, the corresponding nibble of the session key $k^{(i)}$ becomes known. Because in the original permutation case, the two continuous session keys are

$$k^{(i)} = k^{(i)}_{0,L} | k^{(i)}_{0,R} | k^{(i)}_{1,L} | k^{(i)}_{1,R} | \cdots k^{(i)}_{7,L} k^{(i)}_{7,R} | k^{(i)}_{8,L} k^{(i)}_{8,R} \cdots | k^{(i)}_{15,L} | k^{(i)}_{15,R},$$
$$k^{(i+1)} = k^{(i+1)}_{0,L} | k^{(i+1)}_{0,R} | k^{(i+1)}_{1,L} | k^{(i+1)}_{1,R} | \cdots k^{(i+1)}_{7,L} | k^{(i+1)}_{7,R} | k^{(i+1)}_{8,L} | k^{(i+1)}_{8,R} \cdots | k^{(i+1)}_{15,L} | k^{(i+1)}_{15,R}.$$

If Eve detects that the second nibble of ($k^{(i+1)} \oplus k^{(i)}$) is "0000", then she has $k^{(i+1)}_{0,R} = k^{(i)}_{0,R}$. Since $k^{(i+1)}_{0,R}$ is obtained by swapping 0-th and $k^{(i)}_{0,L}$-th elements of $k^{(i)}_R$, if $k^{(i+1)}_{0,R} = k^{(i)}_{0,R}$ then $k^{(i)}_{0,R}$ swaps with itself. It means $k^{(i)}_{0,L} = 0$. From the fourth column of Table 1 Eve knows that the first nibble of ($k^{(i)} \oplus k^{(0)}$) is the $k^{(0)}_{0,L}$. Likewise, if the 18-th nibble of ($k^{(i+1)} \oplus k^{(i)}$) is "0000", then she has $k_{8,R}^{(i+1)} = k_{8,R}^{(i)}$. Since $k_{8,R}^{(i+1)}$ is obtained by swapping 8-th and $k_{8,L}^{(i)}$-th elements of $k_R^{(i)}$, if $k_{8,R}^{(i+1)} = k_{8,R}^{(i)}$ then $k_{8,R}^{(i)}$ swaps with itself. It means $k_{8,L}^{(i)} = 8 = (1000)_2$. From the fourth column of Table 1 Eve knows that xor'ing the 18-th nibble of ($k^{(i)} \oplus k^{(0)}$) with "1000" is the $k_{8,L}^{(0)}$. Gradually all other nibbles of $k^{(0)}$ could be obtained by the attacker in this way.

And in the permutation without out-shuffles situation (see Enhancement I), two consecutive session keys are

$$k^{(i)} = k_{0,L}^{(i)} | k_{0,R}^{(i)} | k_{1,L}^{(i)} | k_{1,R}^{(i)} | \cdots \quad k_{7,L}^{(i)} k_{7,R}^{(i)} | k_{8,L}^{(i)} k_{8,R}^{(i)} \quad \cdots | k_{15,L}^{(i)} | k_{15,R}^{(i)},$$
$$k^{(i+1)} = k_{0,L}^{(i+1)} | k_{1,L}^{(i+1)} | k_{2,L}^{(i+1)} | k_{3,L}^{(i+1)} | \cdots \quad k_{14,L}^{(i+1)} | k_{15,L}^{(i+1)} | k_{0,R}^{(i+1)} | k_{1,R}^{(i+1)} \quad \cdots | k_{14,R}^{(i+1)} | k_{15,R}^{(i+1)}.$$

Now the nibble attack only applies to gain two nibbles of $k^{(0)}$: 17-th and 32-th nibbles, if the first nibble of ($k^{(i+1)} \oplus k^{(i)}$) is "0000", then she has $k_{0,L}^{(i+1)} = k_{0,L}^{(i)}$. Since $k_{0,L}^{(i+1)}$ is obtained by swapping 0-th and $k_{0,R}^{(i)}$-th elements of $k_L^{(i)}$, if $k_{0,L}^{(i+1)} = k_{0,L}^{(i)}$ then $k_{0,L}^{(i)}$ swaps with itself. It means $k_{0,R}^{(i)} = 0$.

From the fourth column of Table 1 Eve knows that the 17-th nibble of (k(i) $\oplus$ k(0)) is the the 17-th nibble of $k^{(0)}$. Similar argument for gaining 16-th nibble of $k^{(0)}$, if $k_{15,R}^{(i+1)} = k_{15,R}^{(i)}$, then $k_{15,L}^{(i)} = 15 = (1111)_2$. So the 16-th nibble of $k^{(0)}$ is the xor'ed result of 16-th nibble of ($k^{(i+1)} \oplus k^{(i)}$) and "1111". All other nibbles of the $k^{(0)}$ will not easily be recovered by observing "0000" nibbles from the last column of the Table 1.

| run | hop | a | b | a $\oplus$ b = c | c(i+1) $\oplus$ c(i) |
|---|---|---|---|---|---|
| i | h0=h(k(i)) | x(i) $\oplus$ k(i) | x(i) $\oplus$ k(0) | k(i) $\oplus$ k(0) | undefined |
| i+h0+1 | h1=h(k(i+h0+1)) | x(i+h0+1) $\oplus$ k(i+h0+1) | x(i+h0+1) $\oplus$ k(0) | k(i+h0+1) $\oplus$ k(0) | k(i+h0+1) $\oplus$ k(i) |
| i=i+h0+1 i+h1+1 | h2=h(k(i+h1+1)) | x(i+h1+2) $\oplus$ k(i+h1+2) | x(i+h1+2) $\oplus$ k(0) | k(i+h1+2) $\oplus$ k(0) | k(i+h1+2) $\oplus$ k(i+1) |
| i=i+h1+1 i+h2+1 | h3=h(k(i+h2+1)) | x(i+h2+3) $\oplus$ k(i+h2+3) | x(i+h2+3) $\oplus$ k(0) | k(i+h2+3) $\oplus$ k(0) | k(i+h2+3) $\oplus$ k(i+2) |
| … | | … | … | … | … |

Table 1. Nibble attack table. In original XOR protocol, hopping offsets $h_0$, $h_1$, $h_2$, … are all 0's, and i does not update. With the hopping the runs, these offsets $h_0$, $h_1$, $h_2$, … are functions of current session keys, e.g., $h_0 = h(k^{(i)})$.

The hopping function is simply defined as a resulting nibble of xor'ing first eight nibbles of the current session key. For instance, the hopping offset $h_0 = h(k^{(i)}) = \bigoplus_{m=0}^{7} k_m^{(i)}$, here $k_m^{(i)}$ is the m-th nibble of the session key $k^{(i)}$. [This hopping_offset formula could be changed to a something like a simple hash.]

With this hopping the runs mechanism equipped in the XOR protocol, even attacker Eve finds "0000" nibble in the last column of the Table 1, she has no way of knowing hopping offsets besides the brute force guessing. Therefore the nibble attack is prevented.

This enhancement makes the nibble attack impossible. Meanwhile it may slow down the calculation speed a little bit, since the next session key is not just one iteration of the permutation, it is (hopping_offset +1) iterations. Note, that "+1" is just to prevent repeat session keys in case of hopping_offset = 0.

### 3.3 Enhancement III – Authenticating Mutually

In general a 3-pass mutual authentication protocol works as follows. Both parties Alice and Bob have a piece of shared secret $k$. Alice initiates the first pass by sending a challenge $F_k(R_A)$, $F_k$ is a kind of encryption (or cryptographic hash) function controlled by $k$, $R_A$ is a random number chosen by Alice. Bob responds with $F_k(R_B)+R_A$ in the second pass, $R_B$ is random number chosen by Bob. In this second pass, Bob is authenticated by Alice, because only Bob is able to extract the random number $R_A$. In the third pass, Alice acknowledges Bob by sending $F_k(R_B+R_A)$ back. In this final pass, Alice is authenticated by Bob, since only Alice is able to restore $R_B$ with their shared secret $k$.

In RFID system, mutual authentication is very important. Without mutual authentication, Reader and Tag could be out of synchronization for the further communication. Because the challenges and responses between Reader and Tag have to keep changing to avoid traceability of the Tag. The XOR protocol is a 2-pass protocol, only Tag is authenticated by Reader and reader is not authenticated by tag. We need to add the third pass to make it a mutual authentication protocol as in Figure 2.

To illustrate we use hopping the runs XOR protocol, we change the next session key index as i+hopping_offset+2 (in stead of "+1" in Enhancement-II) in order to leave a middle permutation for the acknowledge message $c^{(i)}$ of the third pass. It is $c^{(i)} = x^{(i)} \oplus k^{(i+\lfloor (h_i+2)/2 \rfloor)}$, here hi is the hopping_offset, $\lfloor (h_i+2)/2 \rfloor$ takes the greatest integer less than or equal to $(h_i+2)/2$.

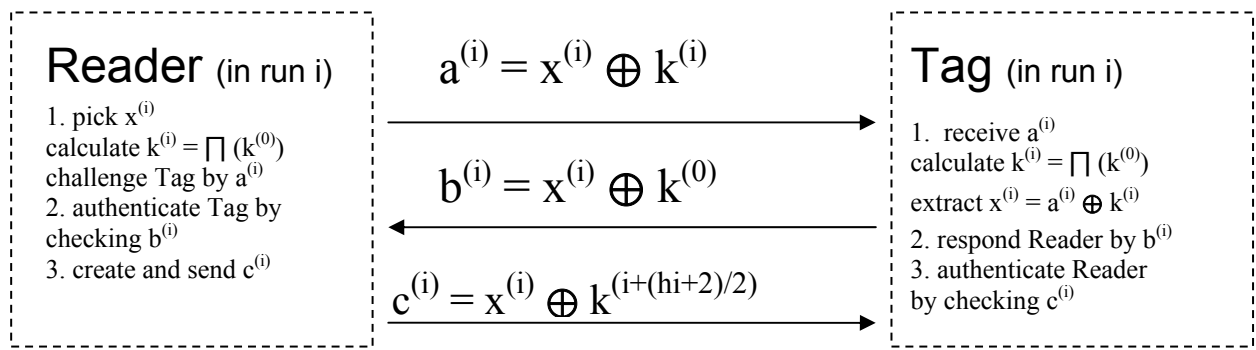

Figure 2. Hopped XOR mutual authentication protocol.

In the first pass, Reader sends a challenge $a^{(i)} = x^{(i)} \oplus k^{(i)}$ to Tag. In the second pass, Tag responds by sending $b^{(i)} = x^{(i)} \oplus k^{(0)}$. Because only legitimate tag is able to extract the challenge $x^{(i)}$ and create the response $b^{(i)}$. By receiving $b^{(i)}$, reader authenticates the tag. In the third pass, reader sends $c^{(i)} = x^{(i)} \oplus k^{(i+\lfloor (h_i+2)/2 \rfloor)}$ back to Tag. Because only legitimate Reader knows $x^{(i)}$ and is able to generate $k^{(i+\lfloor (h_i+2)/2 \rfloor)}$, therefore $c^{(i)}$. After receiving $c^{(i)}$ Tag knows it comes from the right Reader, so Tag authenticates Reader.

## 4 CYCLE COMPARISON EXPERIMENT

In this section, we provide results that compare the cycles of session keys from original VB's XOR algorithm, and the XOR algorithm without that bad shuffle step. Because for a 128-bit key, there are 16 symbols (0 ~ F), the total number of random permutation of 16!*16! is too big to test on PC, we tested two shorter cases: 4-symbol (0 ~ 3) and 8-symbol (0 ~ 7).

For 4-symbol situation, each symbol can be represented in two bits in binary. The key consists of two sets of those 4 symbols, and the key (and session key) length is 2*4*2 bits, i.e., 2 bytes. In ideal case, the number of random permutations for concatenated two sets of 4-symbol is: 4! * 4! = 576.

For 8-symbol situation, each symbol can be represented in 3 bits in binary. The key consists of two sets of those 8 symbols, and the key (and session key) length is 3*8*2 bits, i.e., 6 bytes. In ideal case, the total number of random permutations for concatenated two sets of 8-symbol is: 8! * 8! = 1,625,702,400. The testing results are given in Table 2. Shorter cycles for the XOR without bad shuffle are caused due to too biased distributions of symbols in the initial keys (those weak initial keys), see Appendix A for the detailed explanation and elimination of weak keys.

Table 2. Cycle testing results for two XOR algorithms

|  | 16-bit session key | 48-bit session key |
| --- | --- | --- |
| Number of different initial keys | 1,000 | 100 |
| Number of session keys generated from one initial key | 500 | 10,000 |
| Cycle of the original XOR algorithm | 4 | 32 |
| Cycle of the XOR without bad shuffle | 22 | 9,482 |

## 5 CONCLUSIONS

In this paper we identify the weakness in the XOR authentication protocol proposed by Vajda and Buttyan. We made three enhancements to this protocol: removing bad shuffles, hopping the runs, and authenticating mutually. By these enhancements the XOR protocol is stronger to repeated key attack and nibble attacks proposed by Defend, Fu, and Juels. Our enhancements to the XOR protocol do not introduce extra resource cost. The storage resource needed for the XOR protocol is only 128-bit plus some temporary storage for permutation use, it is extremely light. We believe it's suitable for majority low-cost RFID system application scenarios.

## ACKNOWLEDGEMENTS

We would like to thank Professor Michael Anshel for his encouragement and informative discussions.

Author's E-mail Addresses:
Xiaowen Zhang: zhangx@mail.csi.cuny.edu
Zhanyang Zhang: zhangz@mail.csi.cuny.edu
Xinzhou Wei: xwei@CityTech.Cuny.Edu

# APPENDIX A. 48-BIT SESSION KEY

Generated 100 initial keys are given in Table A-1. Table A-2 and A-3 are cycles for the original XOR and XOR without bad shuffle algorithms.

Table A-1. 100 initial keys

| # |  | # |  | # |  | # |  |
|---|---|---|---|---|---|---|---|
| 00 | 42 42 43 01 71 62 62 37 | 25 | 54 47 00 24 36 53 53 63 | 50 | 51 44 53 65 31 50 57 10 | 75 | 66 45 16 43 74 53 42 43 |
| 01 | 46 40 22 37 10 25 33 57 | 26 | 13 47 15 34 04 37 10 61 | 51 | 06 44 21 14 22 14 54 76 | 76 | 25 45 51 32 54 33 53 25 |
| 02 | 31 43 50 23 60 54 66 04 | 27 | 40 43 50 10 24 31 51 33 | 52 | 50 40 11 11 67 00 43 13 | 77 | 00 47 37 54 36 26 11 14 |
| 03 | 72 41 60 60 20 74 43 71 | 28 | 13 46 61 20 66 00 25 67 | 53 | 43 40 50 63 11 63 41 42 | 78 | 22 43 60 76 10 51 00 05 |
| 04 | 41 41 73 56 45 23 15 72 | 29 | 74 43 31 51 34 15 05 13 | 54 | 00 40 01 14 07 72 02 73 | 79 | 13 43 55 51 46 24 45 34 |
| 05 | 74 40 03 30 24 20 40 44 | 30 | 32 46 07 02 54 37 54 53 | 55 | 47 44 34 27 76 41 77 30 | 80 | 03 41 65 71 67 62 45 10 |
| 06 | 02 42 44 40 61 45 51 46 | 31 | 25 40 34 02 23 32 57 46 | 56 | 43 45 71 22 11 32 55 67 | 81 | 54 45 55 55 13 15 67 47 |
| 07 | 25 47 57 76 47 00 46 25 | 32 | 20 44 02 54 41 73 61 17 | 57 | 74 47 31 41 06 31 34 65 | 82 | 25 41 30 17 61 26 47 34 |
| 08 | 50 41 14 53 37 67 64 63 | 33 | 50 42 62 10 77 24 37 01 | 58 | 12 44 12 05 77 76 41 21 | 83 | 12 47 13 70 15 42 05 13 |
| 09 | 24 40 03 55 31 66 47 21 | 34 | 13 46 57 45 27 72 62 60 | 59 | 54 46 30 32 46 12 74 02 | 84 | 33 43 46 62 62 74 05 42 |
| 10 | 56 41 37 06 37 42 72 43 | 35 | 50 46 71 04 55 24 17 33 | 60 | 26 40 25 77 16 75 52 27 | 85 | 22 42 16 17 13 76 27 33 |
| 11 | 02 43 66 63 50 01 51 04 | 36 | 71 46 34 23 45 06 56 16 | 61 | 65 43 21 55 57 77 34 33 | 86 | 22 41 46 60 72 62 13 14 |
| 12 | 20 42 21 04 32 11 72 06 | 37 | 70 44 13 32 17 54 63 72 | 62 | 76 46 62 56 56 17 15 17 | 87 | 30 46 12 51 07 14 04 62 |
| 13 | 21 45 45 30 32 45 34 23 | 38 | 04 41 13 05 14 22 47 37 | 63 | 53 46 54 64 50 00 15 02 | 88 | 22 42 71 13 75 37 46 53 |
| 14 | 72 44 71 25 65 50 60 07 | 39 | 12 41 50 13 51 50 33 72 | 64 | 32 41 16 43 66 53 73 51 | 89 | 72 46 54 17 07 10 45 13 |
| 15 | 40 40 43 52 12 45 60 20 | 40 | 61 41 45 13 37 55 35 52 | 65 | 47 45 04 23 77 12 05 40 | 90 | 27 41 61 53 52 55 23 60 |
| 16 | 14 43 75 04 62 76 57 30 | 41 | 16 47 44 14 65 12 15 53 | 66 | 05 41 64 30 44 10 10 00 | 91 | 41 40 67 40 10 72 11 35 |
| 17 | 56 46 31 57 33 61 55 31 | 42 | 72 44 20 73 72 07 16 06 | 67 | 60 41 36 77 60 53 45 05 | 92 | 05 41 26 75 42 60 24 33 |
| 18 | 73 45 21 57 43 43 65 27 | 43 | 46 43 12 32 51 72 60 16 | 68 | 06 44 20 46 34 46 76 03 | 93 | 30 42 71 05 10 36 37 34 |
| 19 | 22 46 30 14 03 60 06 46 | 44 | 37 44 70 53 23 63 66 54 | 69 | 33 46 03 40 01 65 55 65 | 94 | 34 41 63 10 31 76 71 14 |
| 20 | 44 41 03 57 20 74 64 60 | 45 | 60 45 15 25 24 57 77 25 | 70 | 01 44 07 70 11 45 75 53 | 95 | 46 43 22 67 01 00 17 56 |
| 21 | 65 46 47 37 44 07 73 72 | 46 | 12 47 51 35 02 51 17 55 | 71 | 67 43 03 75 73 50 16 25 | 96 | 75 47 20 53 62 25 63 12 |
| 22 | 60 45 21 40 50 33 03 11 | 47 | 71 40 17 73 15 03 45 27 | 72 | 73 41 66 55 00 55 26 13 | 97 | 73 42 67 11 11 30 64 51 |
| 23 | 40 41 51 46 40 56 64 42 | 48 | 52 45 46 33 01 22 02 11 | 73 | 22 41 24 41 40 42 45 22 | 98 | 52 44 35 72 53 54 32 12 |
| 24 | 24 46 47 03 54 44 32 73 | 49 | 36 47 52 61 40 17 57 62 | 74 | 34 43 52 22 10 53 16 62 | 99 | 64 47 04 71 04 12 07 47 |

Table A-2. Shortest cycles for XOR without bad shuffle for session keys from the 100 initial keys

| # | cycle | # | cycle | # | cycle | # | cycle | # |  | # |  | # |  | # |  | # |  | # |  |
|---|---|---|---|---|---|---|---|---|---|---|---|---|---|---|---|---|---|---|---|
| 0 | 10000 | 10 | 10000 | 20 | 10000 | 30 | 10000 | 40 | 10000 | 50 | 10000 | 60 | 10000 | 70 | 10000 | 80 | 10000 | 90 | 10000 |
| 1 | 10000 | 11 | 10000 | 21 | 10000 | 31 | 10000 | 41 | 10000 | 51 | 10000 | 61 | 10000 | 71 | 10000 | 81 | 10000 | 91 | 10000 |
| 2 | 10000 | 12 | 3216 | 22 | 10000 | 32 | 10000 | 42 | 10000 | 52 | 10000 | 62 | 10000 | 72 | 10000 | 82 | 10000 | 92 | 10000 |
| 3 | 10000 | 13 | 10000 | 23 | 10000 | 33 | 10000 | 43 | 10000 | 53 | 10000 | 63 | 10000 | 73 | 10000 | 83 | 10000 | 93 | 10000 |
| 4 | 10000 | 14 | 10000 | 24 | 10000 | 34 | 10000 | 44 | 10000 | 54 | 2593 | 64 | 10000 | 74 | 443 | 84 | 10000 | 94 | 10000 |
| 5 | 10000 | 15 | 10000 | 25 | 10000 | 35 | 10000 | 45 | 10000 | 55 | 10000 | 65 | 10000 | 75 | 10000 | 85 | 10000 | 95 | 10000 |
| 6 | 10000 | 16 | 10000 | 26 | 10000 | 36 | 10000 | 46 | 3019 | 56 | 10000 | 66 | 4428 | 76 | 5043 | 86 | 10000 | 96 | 10000 |
| 7 | 10000 | 17 | 10000 | 27 | 10000 | 37 | 10000 | 47 | 10000 | 57 | 10000 | 67 | 10000 | 77 | 10000 | 87 | 10000 | 97 | 10000 |
| 8 | 10000 | 18 | 10000 | 28 | 10000 | 38 | 10000 | 48 | 10000 | 58 | 10000 | 68 | 10000 | 78 | 10000 | 88 | 10000 | 98 | 10000 |
| 9 | 10000 | 19 | 2327 | 29 | 10000 | 39 | 7190 | 49 | 10000 | 59 | 10000 | 69 | 10000 | 79 | 10000 | 89 | 10000 | 99 | 10000 |

Table A-3. Shortest cycles for original XOR algorithm for session keys from the 100 initial keys

| # |  | # |  | # |  | # |  | # |  | # |  | # |  | # |  | # |  | # |  |
|---|---|---|---|---|---|---|---|---|---|---|---|---|---|---|---|---|---|---|---|
| 0 | 36 | 10 | 55 | 20 | 4 | 30 | 18 | 40 | 12 | 50 | 4 | 60 | 5 | 70 | 267 | 80 | 36 | 90 | 2 |
| 1 | 18 | 11 | 6 | 21 | 14 | 31 | 48 | 41 | 33 | 51 | 4 | 61 | 65 | 71 | 6 | 81 | 6 | 91 | 35 |
| 2 | 54 | 12 | 155 | 22 | 6 | 32 | 10 | 42 | 30 | 52 | 7 | 62 | 3 | 72 | 6 | 82 | 42 | 92 | 8 |
| 3 | 127 | 13 | 4 | 23 | 10 | 33 | 2 | 43 | 4 | 53 | 2 | 63 | 4 | 73 | 10 | 83 | 10 | 93 | 16 |
| 4 | 8 | 14 | 414 | 24 | 2 | 34 | 2 | 44 | 6 | 54 | 74 | 64 | 25 | 74 | 130 | 84 | 15 | 94 | 2 |
| 5 | 80 | 15 | 6 | 25 | 3 | 35 | 34 | 45 | 4 | 55 | 24 | 65 | 12 | 75 | 18 | 85 | 2 | 95 | 24 |
| 6 | 6 | 16 | 10 | 26 | 102 | 36 | 18 | 46 | 184 | 56 | 94 | 66 | 20 | 76 | 2 | 86 | 295 | 96 | 6 |
| 7 | 18 | 17 | 10 | 27 | 26 | 37 | 5 | 47 | 9 | 57 | 6 | 67 | 16 | 77 | 6 | 87 | 5 | 97 | 6 |
| 8 | 3 | 18 | 3 | 28 | 6 | 38 | 11 | 48 | 10 | 58 | 8 | 68 | 10 | 78 | 4 | 88 | 48 | 98 | 4 |
| 9 | 28 | 19 | 38 | 29 | 2 | 39 | 6 | 49 | 20 | 59 | 10 | 69 | 6 | 79 | 6 | 89 | 44 | 99 | 32 |

Let's examine those initial keys that yield shorter cycles for XOR without bad shuffle.

Initial key:                          Cycle   Cause of bias
key-12: 20 42 21 04 32 11 72 06       3216    five 2, three 1, three 0 (one 11)
key-19: 22 46 30 14 03 60 06 46       2327    four 0, four 6, three 4 (one 22)
key-39: 12 41 50 13 51 50 33 72       7190    four 1, three 5, three 3 (two 50, one 33)
key-46: 12 47 51 35 02 51 17 55       3019    five 5, four 1 (one 55)
key-54: 00 40 01 14 07 72 02 73       2593    six 0, three 7 (one 00)
key-66: 05 41 64 30 44 10 10 00       4428    six 0, four 4, three 1 (two 10, one 00, one 44)
key-73: 22 41 24 41 40 42 45 22       443     six 2, six 4 (two 22, two 41)
key-76: 25 45 51 32 54 33 53 25       5043    six 5, four 3, three 2 (two 25, one 33)

A balanced initial key should contain all 8 symbols twice. The symbol distribution of the above initial keys are very biased, we can treat these initial keys as weakkeys. In extreme situation, like the key-66, the actual unique symbols are 2, 4, 1, 0, 5, it makes the cycle so short.

We can have a filter to eliminate weakkeys. We did a simple experiment, changing one most frequent symbol to another symbol that was not originally in the initial key, and then all cycles will be maximal to 10000. The following red symbols are changed ones.

key-12: 20 45 21 04 32 11 72 06
key-19: 22 47 30 14 03 60 06 46
key-39: 12 46 50 13 51 50 33 72
key-46: 12 47 61 35 02 51 17 55
key-54: 00 45 01 14 07 72 02 73
key-66: 75 41 64 30 44 10 10 00
key-73: 22 41 74 41 40 42 45 22
key-76: 25 45 71 32 54 33 53 25

From these experiments, we can see that if symbol distribution in a initial key is not too biased, then through the enhanced XOR algorithm we can generate enough number of session keys for the subsequent authentication uses.